\documentclass[conference,a4paper]{IEEEtran}

\usepackage{amssymb}
\usepackage{amsmath}
\newcommand{\F}{\mathbb F}
\usepackage[utf8]{inputenc}
\usepackage{color}
\usepackage{xspace}

\newcommand{\refeq}[1]{Eq.~\ref{#1}}
\newcommand{\set}[1]{\left\{#1\right\}}
\newcommand{\size}[1]{\left|#1\right|}
\newcommand{\intervalle}[2]{\left\{#1,\dots,#2\right\}}
\newcommand{\Min}{\operatorname{Min}}
\newcommand{\etal}{\emph{et al.}\xspace}

\newtheorem{theorem}{Theorem}

\newtheorem{proposition}{Proposition}
\newtheorem{corollary}{Corollary}

\title{Exhaustive Search for Small Dimension Recursive\\MDS Diffusion Layers for Block Ciphers\\and Hash Functions}
\author{\IEEEauthorblockN{Daniel Augot}\IEEEauthorblockA{INRIA Saclay -- Île-de-France \&\\ LIX -- École Polytechnique\vspace{-5mm}}
\and
\IEEEauthorblockN{Matthieu Finiasz}\IEEEauthorblockA{CryptoExperts}}

\begin{document}
\maketitle

\begin{abstract}\boldmath
  This article presents a new algorithm to find MDS matrices that are well
  suited for use as a diffusion layer in lightweight block ciphers. Using an
  recursive construction, it is possible to obtain matrices with a very compact
  description. Classical field multiplications can also be replaced by simple
  $\F_2$-linear transformations (combinations of XORs and shifts) which are much
  lighter. Using this algorithm, it was possible to design a $16\times 16$ 
  matrix on a 5-bit alphabet, yielding an efficient 80-bit diffusion layer with
  maximal branch number.
\end{abstract}


\section{Introduction}

There are many ways to construct Maximum Distance Separable (MDS) codes and,
depending on the target application, some are better than others. One
application of MDS codes is the design of linear diffusion layers in block
ciphers (or cryptographic hash functions). The linear code consisting of words
formed by the concatenation of inputs and outputs of a linear diffusion layer
should have the best possible minimum distance to ensure optimal diffusion.
Hence, MDS codes having the largest possible minimal distance, they are a good
choice from a security point of view. However, they also require a dense matrix
to be used, leading to a large description and somewhat slow evaluation.

In 2011, Guo~\etal introduced the LED block cipher~\cite{led} and the PHOTON
hash function family~\cite{photon} where they use a $4\times 4$ MDS diffusion
matrix constructed as a power of a companion matrix. LED and PHOTON being
lightweight designs, this structure allows for a much more compact description
of the diffusion layer, which is crucial in this context. In the beginning of
2012, Sajadieh~\etal~\cite{Sajadieh:FSE2012} used a similar construction, and
presented it as a Feistel-like recursive construction to design efficient and
compact (in terms of code size or gate usage) MDS diffusion matrices for block
ciphers. In addition, they also replaced the finite field operations present in
PHOTON by simpler $\F_2$-linear operations, again improving the efficiency and
compactness of the construction.  The same design strategy was then used by
Wu~\etal~\cite{WuWangWu:SAC2012} to obtain optimal diffusion layers using the
smallest possible number of XOR gates when specifically targeting hardware
implementation. These constructions are probably not the best choice for a
software implementation running on a computer, but they are perfect for
lightweight designs where MDS diffusion was usually not considered an option.

In this article, we continue this idea and propose an algorithm to build even
larger MDS matrices, using the same recursive construction. Our main target is
to obtain full state-wide optimal diffusion, as opposed to designs like the AES
where the MDS diffusion is only applied to a small part of the state, offering
optimal diffusion in this part, but sub-optimal diffusion in the state as a
whole.

This article is structured as followed. After presenting the notations we will
use throughout the paper, we start by recalling the works of Sajadieh~\etal and
of Wu~\etal and the results they were able to obtain. We then present the new
approach we used for our construction and a series of theoretical results
supporting it. Eventually, we expose the results we obtained for $8\times 8$
diffusion matrices with symbols of 4 bits and $16\times 16$ matrices with
symbols of $5$ bits.

\subsection{Notation}

The final result we aim at is a square matrix operating on $\ell$ symbols in
$\F_{q^s}$.  For any ring $R$, we denote $M_\ell(R)$ the set of $\ell\times \ell$
matrices with coefficients in $R$. We will term \emph{symbolic} a polynomial
$p(X)\in\F_{q}[X]$ or $\F_{q^s}[X]$, or an $\ell\times \ell$ square polynomial
matrix $M(X)\in M_\ell(\F_q[X])$, where $X$ is an \emph{indeterminate}. Given
\emph{values} $\alpha$ $\in \F_{q^s}$, or $L$ an $s\times s$ square matrix in $\F_q$, we
will get \emph{values} by substituting $\alpha$ or $L$ in the \emph{symbolic}
polynomial $p(X)$, or in the \emph{symbolic} matrix $M(X)$. Using the standard
computer science notation, we will denote such substitutions
\[
\begin{array}{ll}
p(X\gets a)\in\F_{q^s},
&\quad p(X\gets L)\in M_\ell(\F_q)\\
M(X\gets a)\in M_\ell(\F_{q^s}),
&\quad M(X\gets L)\in M_\ell(M_s(\F_q))
\end{array}
\]
The aim is to find an $\ell\times\ell$ symbolic matrix $M(X)$ and an
$s\times s$  matrix $L$ in $\F_q$ such that the mapping  $M_L:=M(X\gets L)$:
\[
M_L:\begin{array}[t]{ccl}
\left(\F_{q^s}\right)^\ell&\rightarrow&\left(\F_{q^s}\right)^\ell\\
v&\mapsto &M_L\cdot v
\end{array}
\]
has  \emph{maximum branch number} $\ell+1$, i.e.\ such that, denoting $w(v)$ the Hamming weight of $v$:
\[
\min_{v\neq0}\set{w(v)+w(M_L\cdot v)}=\ell+1.
\]

Let $C_{M_L}$ be the code of length $2\ell$ over the alphabet $\F_{q^s}$, whose
codewords are $\left(v||M_L\cdot v\right)$ for all $v\in
\left(\F_{q^s}\right)^\ell$, or equivalently, whose generating matrix is
\[
\begin{bmatrix}
\,I \, |\, M_L
\end{bmatrix}
\]
Requiring that $M_L$ has maximum branch number is equivalent to asking
that the code $C_{M_L}$ has \emph{minimum distance} $\ell+1$.  Note
that, $L$ being $\F_q$-linear, $C_{M_L}$ is linear over $\F_q$, but
\emph{non-linear} over the field $\F_{q^s}$.  Still, the Singleton
bound also holds for non linear codes.
\begin{theorem}[Singleton bound\cite{Singleton:IT64}]
  Consider a $Q$-ary unspecified alphabet. Let $C$ be a $Q$-ary code of
  length $n$, minimal distance $d$. Then $\size C\leq Q^{n-d+1}$.
\end{theorem}

A code is called Maximum Distance Separable (MDS) if the equality
$d=n-k+1$ holds. In our case $n=2\ell$ and $k=\ell$, so being MDS
requires $d=\ell +1$. We briefly recall the MDS conjecture for linear
codes, when $Q$ is a prime power, the alphabet is given a field
structure, and the code is linear: except for particular or degenerate
cases, if there exists a $Q$-ary linear MDS code of length $n$, then
$n\leq Q+1$. Which translates to $2\ell \leq q^s +1$. The mentioned
degenerate cases above do not cover the cases $n=2\ell$, $k=\ell$ we
are interested in.

Given an $s\times s$ matrix $L$ in $\F_q$, we denote $\Min_L(X)$ the
minimal polynomial of $L$, which is the smallest degree non zero
polynomial $p(X)\in\F_q[X]$ such that $p(X\gets L)=0$.

\section{Previous Work}

Sajadieh, Dakhilalian, Mala and Sepehrdad~\cite{Sajadieh:FSE2012} describe their
diffusion layer as a Feistel-like recursive structure.  One application of such
a diffusion layer to an $\ell$ symbol state $S_0$ consists in $\ell$ successive
applications of a sub-layer which adds to one symbol a linear combination of the
others, and circularly shifts the state (as in a generalised Feistel cipher).
This sub-layer can be represented as the multiplication by a companion matrix
$C$:\vspace{-3mm}
$$S_{i+1} = C\cdot S_i, \textrm{ with } C = \begin{bmatrix}
  0 & 1 & 0 & \dots & 0\\
  0 & 0 & 1 & \dots & 0\\
  \vdots & & \ddots& \ddots & \vdots\\
  0 & 0 & \dots & 0 & 1\\
  1 & c_1 & c_2 & \dots & c_{\ell-1}\end{bmatrix}.$$ 
Where each $c_i$ represents an $\F_q$-linear transformation and a 1 the identity
function. $c_0$ is chosen equal to 1, making the inverse transformation almost identical. The output $S_\ell$ of the full diffusion layer can thus be expressed
as $S_\ell = C^\ell\cdot S_0$.

Before presenting the construction they propose, we need to recall a few facts
about MDS codes over rings, as the matrices $M$ and $C$ no longer have
coefficients in a field.

\subsection{MDS Codes over Commutative Rings}

Over a finite field, the criterion for an $\ell\times\ell$ matrix to have maximum
branch number (and define an MDS code) is that all its minors (of any size $\leq
\ell$) should be non-zero. Over a commutative ring, this criterion simply
translates to minors being \emph{invertible}.

\begin{proposition}
  Let $M\in M_\ell(R)$ be a matrix over a finite commutative ring $R$. Then $M$
  has maximum branch number if and only if all the minors of $M$ are invertible
  in $R$.
\end{proposition}
\begin{IEEEproof} For any square matrix $A$ over a commutative ring, denoting
  $A'$ the adjugate matrix of  $A$, we have the following relation:
\begin{equation}\label{eq.tcomatrix}
A'\cdot A =A\cdot A'=\det(A)\cdot I
\end{equation}
We first prove that if $M$ is non MDS, then there is a minor of $M$
which is not invertible. Since $M$ is not MDS and does not have
maximum branch number, there exists $v\in R^\ell$, of weight say $w$,
such that $M\cdot v$ has weight strictly less than $\ell+1-w $, i.e.\
it has at least $w$ zero coordinates.  Consider
$J=\{j_1,...,j_w\}\subset\intervalle 1\ell$ a set of non zero
coordinates of $v$ and $I=\{i_1,...,i_w\}\subset \intervalle 1\ell$ a
subset of the zero coordinates of $M\cdot v$.  Then the submatrix
$M_{|I,J}$ sends $v_{|J}$ to 0.  Using~\refeq{eq.tcomatrix}, we get\vspace{-2mm}
\[
\det (M_{|I,J})\cdot v=0,\vspace{-1mm}
\]
meaning that the $\det(M_{|I,J})$ is non invertible, since $v$ has at least one
non zero coordinate.

Suppose now that $M$ has maximum branch number, then for any integer $w$ and any
$v$ of weight $w$, with non zero positions located in $I$, $\size I=w$, we have
$w(M\cdot v)\geq \ell+1-w$.  In particular for any subset $J\subset\intervalle
1\ell $ of size $w$, $\left(M\cdot v\right)|{_J}\neq0$.  Considering the matrix
$M_{|I,J}$, the mapping $x\mapsto M_{|I,J} \cdot x$ thus has a kernel equal to
$\set 0$. It is thus invertible, and using~\refeq{eq.tcomatrix}, $\det(
M_{|I,J})$ is invertible in $R$.
\end{IEEEproof}

\subsection{The Method}

Going back to the Sajadieh~\etal construction, they propose to choose an $s\times
s$ binary matrix $L$ and have each $c_i$ be a non zero polynomial $c_i(X) \in
F_2[X]$ evaluated in $L$. In practice, they restrict to $c_i(X)$ of degree 1 or
2. Wu~\etal~\cite{WuWangWu:SAC2012} use the same construction but restrict to
$c_i(X)$ which are monomials in $L$.  Restricting to such polynomials in $L$
makes products of $c_i$ commutative.  Thanks to the previous proposition, the
search for an efficient diffusion matrix with maximum branch number can be done
in two steps:
\begin{enumerate}
\item\label{item.searchM} exhaustively search for a good symbolic matrix $M(X) =
  C(X)^\ell$ (where $C(X)$ is an $\ell\times\ell$ symbolic companion matrix).
  Here, good means that the set $m(X)$ of all the minors of $M(X)$ does not
  contain the null polynomial.
\item\label{item.seachL} find a suitable $\F_2$-linear operator
  $L:\F_2^s\rightarrow\F_2^s$, such that all the matrices in $m(L)$ (the set
  obtained when applying $X\gets L$ to the minors in $m(X)$) are invertible.
\end{enumerate}

The first part of the search outputs a set of symbolic matrices, each one with a
set of constraints associated to it.  Each of these constraints is a polynomial
that has to be invertible when evaluated in $L$, so fewer distinct polynomials
is usually better. Sajadieh~\etal also focus on having low degree polynomials:
this way, picking an $L$ matrix that has a minimal polynomial $\Min_L(X)$
irreducible and of higher degree than all the minors in $m(X)$ will always give
an MDS matrix. Wu~\etal rather focus on some specific matrices $L$ with a single
XOR gate and directly check whether they verify each constraint.

\subsection{Obtained Results}

Using the previous method, Sajadieh~\etal were able to exhaustively search all
$M(X) = C(X)^\ell$ with polynomials $c_i(X)$ of degree 1 for values of $\ell$ up
to 8. No results were found for $\ell > 4$ and only very few solutions exist for
$\ell =2$, 3 or 4, with different sets of constraints.  They propose various
matrices $L$ verifying these constraints for sizes $s$ ranging from 4 bits to 64
bits. They also present some solutions for $C$ when $\ell \in \set{5,6,7,8}$
using polynomials $c_i$ of degree 2.

We were able to run the full exhaustive search for $\ell = 8$ with polynomials
$c_i$ of degree $2$ looking for constraints $m(X)$ of the smallest possible
degree. We found 12 solutions where all minors can be decomposed into factors of
degree at most 14. It is thus possible to use any of these 12 solutions with a
matrix $L$ having a minimal polynomial $\Min_L(X)$ of degree 15 or more and get
a matrix with maximum branch number. However, a $\Min_L(X)$ of degree 15
requires $L$ to operate on at least $s = 15$ bits. This is much more than the
MDS bound which implies that the number of possible symbols $2^s$ should be at
least $2\ell$. For $\ell = 8$, symbols of $s = 4$ bits could be possible.

Indeed, as shown by Wu~\etal, symbols of $s=4$ bits are possible. Instead of
searching for $m(X)$ with only low degree factors, they search for $m(X)$
containing no multiples of an irreducible polynomial $p(X)$ of given degree.
This way, any $L$ such that $\Min_L(X) = p(X)$ will yield an $M(L)$ with maximum
branch number. Using $p(X) = X^4 + X + 1$, they found the companion
matrices listed in Table~\ref{tab.results_wu}.
\begin{table}\centering
$$\begin{array}{|c|c|}\hline
\ell = 5 & [1, L^2, L^{-1}, L^{-1}, L^2] \\\hline
\ell = 6 & [1, L^{-2}, L^{-1}, L^2, L^{-1}, L^{-2}] \\\hline
\ell = 7 & [1, L, L^{-5}, 1,1, L^{-5}, L] \\\hline
\ell = 8 & [1, L^{-3}, L, L^3, L^2, L^3, L, L^{-3}] \\\hline
\end{array}$$
\caption{\label{tab.results_wu}Companion matrices found by Wu~\etal~\cite{WuWangWu:SAC2012}}\vspace{-5mm}
\end{table}

\section{New Approach}

\subsection{Changing the Ordering of the Searches}

The first thing we noted when performing our experiments on $8\times 8$ matrices
using degree 2 polynomials $c_i$ is that the symbolic computation of minors are
very expensive, especially as the obtained polynomials are of rather high degree
(even if they decompose in small factors). In order to make the search more
efficient, and thus be able to explore larger parameters, we needed to get rid
of symbolic computation. This is what we did by reordering the search:

\begin{enumerate}
\item\label{item.seachL2} instead of considering any $\F_2$-linear operator $L$,
  focus only on operators having a given minimal polynomial $\Min_L(X)$ of degree
  $d$,
\item\label{item.searchM2} exhaustively search for a good symbolic matrix
  $M(X)\in\F_2[X]$ whose symbolic minors are all invertible under $X\gets L$.
\end{enumerate}

The crucial remark is that requiring that some minor $m(X)$ is invertible under
$m(X\gets L)$ is the same as requiring that $m(X)$ is invertible in
$\F_2[X]/\Min_L(X)$. The minors can thus all be computed directly in
$\F_2[X]/\Min_L(X)$. In particular, when $\Min_L(X)$ is irreducible,
$\F_2[X]/\Min_L(X)$ is a field, and all computations can be done directly in
$\F_{2^d}$.

If we define $\alpha$ as a root of $\Min_L(X)$ in $\F_{2^d}$, requiring that
$M(X\gets L)$ has maximal branch number is the same as requiring that
$M(X\gets\alpha)$ has maximal branch number in the classical sense. This paves
the way to theorems, see below, and to much faster computations, enabling
exhaustive searches which were out of reach.

\subsection{Case of General Symbolic Matrices}
We state the theorems for arbitrary finite fields, but our primary target
remains $\F_2$ only.
\begin{proposition}\label{prop.field}
  Let $\F_q$ be a finite field. Let $M(X)$ be an $\ell\times \ell$
  matrix with coefficients in $\F_q[X]$. Let $L$ be an $s\times s$
  matrix with coefficients in $\F_q$, such that its minimal polynomial
  $p(X) = \Min_L(X)$ is irreducible of degree $d$. Then $M(X\gets L)$
  has maximum branch number if and only if $M(X\gets\alpha)$ is an MDS
  matrix over the field
\[
\F_q[\alpha]=\F_q[X]/p(X)\simeq\F_{q^d}
\]
where $\alpha$ is a root of $\Min_L(X)$ in $\F_{q^d}$.
\end{proposition}
\begin{IEEEproof}
  Matrix $M(X\gets L)$ has maximal branch number if and only if all
  the minors of all sizes of $M(X)$ are invertible after $X\gets L$,
  which is the same as saying that they are co-prime with $\Min_L(X) =
  p(X)$. Since $p(X)$ is irreducible, minors of $M(X)$ should simply
  not be multiples of $p(X)$, which is equivalent to saying that
  $\alpha\in \F_{q^d}$ should not be a root of any minor. This will be
  the case if and only if $M(X\gets\alpha)$ is an MDS matrix over
  $\F_{q^d}$.
\end{IEEEproof}
Assuming that the MDS conjecture holds true, we get directly a bound
on $\ell$ in terms of $s$. Since we are dealing with codes of even
length $2\ell$, the conjecture gives  $2\ell\leq Q$ where $Q$ is the
size of the field of symbols.
\begin{corollary}\label{coro.MDSbound}
  Suppose that the minimal polynomial of $L$ is irreducible of degree $d$, then
  $\ell\leq \frac12 q^{d}$ is a necessary condition for $M(X\gets L)$ to be MDS.
  Also, if $L$ operates on elements of $\F_q^s$, the degree of its minimal
  polynomial is at most $s$, so $d\leq s$. In the case $q=2$, we have the bound
  $s\geq 1+\lceil\log_2(\ell)\rceil$.
\end{corollary}

Next proposition shows that for the case of a matrix $L$ with
irreducible minimal polynomial $p(X)$, the computation needs to be
done only once, and that will encompass all matrices $L$ with any
irreducible minimal polynomial $p(X)$ of the same degree $d$, since
all finite fields of the same extension degree over the ground field
are isomorphic.

\begin{proposition}\label{prop.equivalence}
  Consider $p_1(X),p_2(X)\in\F_q[X]$ two irreducible polynomials of same degree $d$. Consider
  $\alpha_1\in\F_q[X]/p_1(X)$, and $\alpha_2\in\F_q[X]/p_2(X)$ such that
  $p_1(\alpha_1)=p_2(\alpha_2)=0$. Now choose $\alpha'_1\in\F_q[X]/p_2(X)$ such that
  $p_1(\alpha'_1) = 0$. There exists a polynomial $K(X) \in \F_q[X]$ such that $\alpha'_1 =
  K(\alpha_2)$. We have an $\F_q$-isomorphism:
\[
\sigma:\begin{array}[t]{lcl}
 \F_q[X]/p_1(X)&\rightarrow &\F_q[X]/p_2(X)\\
 \alpha_1&\mapsto &\alpha'_1
\end{array}
\]
Suppose that $M_1(X)$ is such that $M_1(X\gets\alpha_1)$ is MDS. Then, any matrix $M_2(X)$ such
that $M_2(X\gets\alpha_2)=\sigma(M_1(X\gets \alpha_1))$ is such that $M_2(X\gets\alpha_2)$ is
MDS. Any $\F_q$-linear operator $L$ with minimal polynomial $\Min_L(X) = p_2(X)$ will then
define a matrix $M_2(X\gets L)$ with maximum branch number. Such a matrix $M_2(X)$ will have the
same recursive structure as $M_1(X)$ and can be computed as:
$$M_2(X) = M_1(X \gets K(X)).$$

\end{proposition}
\begin{IEEEproof}
  Elements $\alpha_1$ and $\alpha'_1$ have the same minimal polynomial, so there exists a field
  isomorphism $\sigma$ sending one onto the other. This field isomorphism preserves the
  invertibility of matrices and also preserves the MDS property of a matrix, so
  $\sigma(M_1(X\gets \alpha_1))$ is MDS over the field $\F_q[X]/p_2(X)$.

  If $M_2(X\gets \alpha_2) = \sigma(M_1(X\gets \alpha_1))$ it is MDS, and we know from
  Proposition~\ref{prop.field} that any $L$ with the same minimal polynomial $p_2(X)$ as
  $\alpha_2$ will define a matrix $M_2(X\gets L)$ with maximum branch number.

  The field isomorphism $\sigma$ commutes with any polynomial operation, so $\sigma(M_1(X\gets
  \alpha_1)) = M_1(X\gets \alpha'_1)$. So $M_2(X \gets \alpha_2) = M_1(X\gets K(\alpha_2))$.
  Substituting $X$ by $K(X)$ in $M_1$ thus gives a valid matrix $M_2(X)$.

  Concerning the recursive structure of $M_1$ and $M_2$, if $M_1(X) = C_1(X)^\ell$ one can
  define $C_2(X) = C_1(X\gets K(X))$ and matrix $M_2(X) = C_2(X)^\ell$ to get $M_2(X\gets
  \alpha_2) = \sigma(M_1(X\gets \alpha_1))$ as expected.

  Finally, matrix $M_1(X)$ will usually be computed as $C_1(X)^\ell\bmod p_1(X)$ so as to keep
  the degree of the polynomials bounded. As $M_1(X)$ is only evaluated on elements of minimal
  polynomial $p_1(X)$ this does not change anything. This reduction modulo $p_1(X)$ is
  compatible with the substitution $X\gets K(X)$ if matrix $M_2(X)$ is reduced modulo $p_2(X)$:
  $M_2(X) = M_1(X\gets K(X)) \bmod p_2(X)$. This comes from the equality $p_1(K(X)) = 0 \bmod
  p_2(X)$, which in turn is true because $\alpha_2$ is a root of $p_1(K(X))$.
\end{IEEEproof}

As we have seen, when $\Min_L(X)$ is irreducible, everything happens exactly as
in the finite field MDS case. The following results show that when $\Min_L(X)$
is not irreducible, similar results can apply.

\begin{corollary}\label{coro.CRT}
  Let $\F_q$ be a finite field. Let $M$ be a $\ell\times \ell$ matrix with
  coefficients in $\F_q[X]$. Let $L$ be a $s\times s$ matrix with coefficients
  in $\F_q$, with minimal polynomial $p(X)=p_1(X)p_2(X)$, where $p_1(X)$ and
  $p_2(X)$ are co-prime.  Then $M(X\gets L)$ has maximum branch number if and
  only if both matrices
\[
M_1=M(X\gets\alpha_1),\quad M_2=M(X\gets\alpha_2)
\]
have maximum branch number  over the rings
\[
\F_q[\alpha_1]=\F_q[X]/p_1(X),\quad
\F_q[\alpha_1]=\F_q[X]/p_2(X)
\]
where $\alpha_1= X\bmod p_1(X)$, and $\alpha_2= X\bmod p_2(X)$.
\end{corollary}
\begin{IEEEproof}
  From the Chinese Remainder Theorem, a minor $m(X)$ is invertible $\bmod\ p(X)$
  if and only if it is invertible both $\bmod\ p_1(X)$ and $\bmod\ p_2(X)$.
  Extending this to all minors implies that $M_1$ and $M_2$ have maximum branch
  number.
\end{IEEEproof}

Next we study the case $\Min_L(X)$ is a power of an irreducible polynomial.
\begin{corollary}\label{coro.power}
  Let $\F_q$ be a finite field. Let $M$ be a $\ell\times \ell$ matrix with
  coefficients in $\F_q[X]$. Let $L$ be a $s\times s$ matrix with coefficients
  in $\F_q$, with minimal polynomial $p(X)=p_1(X)^{e}$, where $p_1(X)$ is
  irreducible. Then $M(X\gets L)$ has maximal branch number if and only if
\[
M_1=M(X\gets\alpha)
\]
is an MDS matrix over the field
\[
\F_q[\alpha]=\F_q[X]/p_1(X)
\]
where $\alpha= X\bmod p_1(X)$.
\end{corollary}
\begin{IEEEproof}
  As in the proof of Corollary~\ref{coro.CRT}, it is enough to remark that a
  minor $m(X)$ is invertible $\bmod\ p_1(X)^e$ if and only it is non zero
  $\bmod\ p_1(X)$.
\end{IEEEproof}

We can now extend these results to the most general case for $\Min_L(X)$.

\begin{corollary}\label{coro.generalcase}
  Let $\F_q$ be a finite field. Let $M$ be an $\ell\times \ell$ matrix
  with coefficients in $\F_q[X]$.  Let $L$ be a $s\times s$ matrix
  with coefficients in $\F_q$, with minimal polynomial
  \[p(X)=p_1(X)^{e_1}\dots p_k(X)^{e_k},
  \] where $p_1(X),\dots,p_k(X)$ are irreducible. Then $M(X\gets L)$
  is MDS if and only if $\forall i \leq k, M(X\gets \alpha_i)$ is MDS
  over $\F_q[\alpha_i]$ (with $\alpha_i = X\bmod p_i(X)$).  As a
  consequence, $\ell \leq \frac{q^d}2$, where $d=\min \{\deg p_i(X),
  i=1,\dots,k\}$.
\end{corollary}
\begin{IEEEproof}
From Corollaries~\ref{coro.MDSbound}, \ref{coro.CRT}, and~\ref{coro.power}. 
\end{IEEEproof}

These theorems clearly indicate that to get a large maximal diffusion matrix, it
is always preferable to use an operator $L$ with irreducible minimal polynomial.

\section{Experimental Results}

Even though all the results from the previous section hold over $\F_q$, our
primary focus being lightweight block ciphers, we only ran experiments on
extensions of $\F_2$.

\subsection{Operating on $4$ bit Blocks}

To test our algorithm, our first targets were the results of
Wu~\etal~\cite{WuWangWu:SAC2012}. We wanted to go through all recursive matrices
with $\ell=8$ and $s=4$ and see how many $8\times 8$ MDS matrices we could
obtain.  There are 3 irreducible polynomials of degree 4 on $\F_2$, but as
stated in Proposition~\ref{prop.equivalence}, performing the search for only one
of them is enough. We chose the minimal polynomial $p(X) = X^4 +X + 1$. We then
simply ran an exhaustive search through the $15^7$ companion matrices $C$ with
$c_0 = 1$ and coefficients $c_i$ in $\F_{2^4}^*$. For each of these $C$ we
computed $M = C^8$ and checked whether all the minors of $M$ (computed in
$\F_{2^4}$) were non zero. Noting $\alpha\in\F_{2^4}$ a root of $p(X)$, we found
the following solutions for $[c_0,\dots,c_7]$:

$$\arraycolsep=2pt\begin{array}{cllllllllll}
S_0 = & [ 1,& \alpha^{3},& \alpha^{4},& \alpha^{12},& \alpha^{8},& \alpha^{12},& \alpha^{4},& \alpha^{3} &]\\
S_1 = & [ 1,& \alpha^{6},& \alpha^{8},& \alpha^{9},& \alpha,& \alpha^{9},& \alpha^{8},& \alpha^{6} &]\\
S_2 = & [ 1,& \alpha^{12},& \alpha,& \alpha^{3},& \alpha^{2},& \alpha^{3},& \alpha,& \alpha^{12} &]\\
S_3 = & [ 1,& \alpha^{9},& \alpha^{2},& \alpha^{6},& \alpha^{4},& \alpha^{6},& \alpha^{2},& \alpha^{9} &]\\\noalign{\vskip1mm}
S_4 = & [ 1,& \alpha^{7},& \alpha^{2},& \alpha^{11},& \alpha^{13},& \alpha^{11},& \alpha^{2},& \alpha^{7} &]\\
S_5 = & [ 1,& \alpha^{14},& \alpha^{4},& \alpha^{7},& \alpha^{11},& \alpha^{7},& \alpha^{4},& \alpha^{14} &]\\
S_6 = & [ 1,& \alpha^{13},& \alpha^{8},& \alpha^{14},& \alpha^{7},& \alpha^{14},& \alpha^{8},& \alpha^{13} &]\\
S_7 = & [ 1,& \alpha^{11},& \alpha,& \alpha^{13},& \alpha^{14},& \alpha^{13},& \alpha,& \alpha^{11} &]\\
\end{array}$$

The solution presented by Wu~\etal (last line of Table~\ref{tab.results_wu})
corresponds to our solution $S_2$. The mapping $\alpha \rightarrow \alpha^2$ is
a morphism in $\F_{2^4}$, so it transforms an MDS matrix into another MDS
matrix. It allows to group the 8 solutions in two classes of equivalent
solutions $\set{S_0,S_1,S_2,S_3}$ and $\set{S_4,S_5,S_6,S_7}$.

The whole exhaustive search (about $2^{27.3}$ matrices) took 2 days on a single
core using Magma~\cite{magma}. The same computation using symbolic polynomials
would probably have taken a few months.

Any of these solution can then be used with a matrix $L$ having minimal
polynomial $\Min_L(X) = X^4+X+1$. This $L$ matrix can be a binary $4\times4$
matrix, yielding a 32 bit diffusion layer, but an $8\times 8$ matrix with
suitable minimal polynomial can also be used to obtain a 64 bit diffusion layer.

\medskip One thing that can be noted about these solutions is their symmetry: in
every solution, $c_1 = c_7$, $c_2 = c_6$, and $c_3 = c_5$. The same symmetry can
be observed in the solutions found by Wu~\etal (see Table~\ref{tab.results_wu}).
We also ran the full exhaustive search for $\ell = 4$ and $s=3$ and found a
single class of solutions (3 equivalent solutions) which is also symmetric $[ 1,
\alpha^3, \alpha, \alpha^3 ]$. For $\ell = 5$ and $s=4$, we found 60 solutions
in 15 classes among which only 3 classes are symmetric. For $\ell = 6$ and
$s=4$, we found 36 solutions in 9 classes among which only 3 classes are
symmetric. This symmetry can thus be observed both for odd and even values of
$\ell$. One thing we proved is that for a given MDS solution, its symmetric is
also MDS.  This is due to the fact that the inverse of a recursive layer can be
expressed as a recursive layer using the symmetric coefficients, and the inverse
of an MDS matrix is also MDS. However, this property does not explain why
symmetric solutions exist, or why when the MDS bound is ``tight'', that is, when
$2\ell = 2^s$, only symmetric solutions exist.

\subsection{Operating on $5$ bit Blocks}

With symbols of $s=5$ bits, the MDS conjecture indicates that matrices of size
$16\times 16$ can have maximal branch number. There are 6 irreducible
polynomials of degree 5 over $\F_2$ and we used Magma's choice $p(X) =
X^5+X^2+1$ for our search.

In our previous experiment with $s=4$, the full exhaustive search required to
test $15^7 \simeq 2^{27.3}$ matrices. Here, the full search would require to
test $31^{15} \simeq 2^{74.3}$ matrices, which is not feasible. However, all
solutions we found for small instances seem to display a nice symmetry. Instead
of exhaustively searching all companion matrices, we only search ``symmetric''
companion matrices. This reduces the number of matrices to test to $31^8\simeq
2^{39.6}$. The number of tests can be reduced even further by taking into
account the equivalence classes with respect to $\alpha \rightarrow \alpha^2$.
For $c_9$ (the middle coefficient) we only tested the values $\set{1, \alpha,
  \alpha^3, \alpha^5, \alpha^7, \alpha^{11}, \alpha^{15}}$, reducing the tests
to $7\cdot 31^7\simeq 2^{37.4}$.

Testing if a $16\times 16$ matrix is MDS also requires to compute much more
minors than for an $8\times 8$ matrix. However, most candidate matrices possess
null coefficients, or null minors of size 2, and computations can be interrupted
early. The total exhaustive search took about 80 days of CPU time, and was run
in about two weeks on a single quad-core computer. In the end, only two
solutions were found (each one representing a class of 5 solutions):
$$\arraycolsep=2pt\begin{array}{cllllllllll}
S_0 = & [ 1,& \alpha^{17},& \alpha,& \alpha^{9},& \alpha^{12},& \alpha,& \alpha^{27},& \alpha^{25},& \alpha^{7},&\\
&&&& \alpha^{25},& \alpha^{27},& \alpha,& \alpha^{12},& \alpha^{9},& \alpha,& \alpha^{17} ] \\\noalign{\vskip1mm}
S_1 = & [ 1,& \alpha^{20},& \alpha^{25},& \alpha^{3},& \alpha^{27},& \alpha^{19},& \alpha^{9},& \alpha^{27},& \alpha^{15},&\\
&&&& \alpha^{27},& \alpha^{9},& \alpha^{19},& \alpha^{27},& \alpha^{3},& \alpha^{25},& \alpha^{20} ] \\
\end{array}$$

There might be other ``non-symmetric'' solutions as we have not searched for
them, but we conjecture there are none.  Each of the solutions we found can be
used with a $5\times 5$ binary matrix $L$ with $\Min_L(X) = X^5+X^2+1$ to obtain
an optimal 80 bit diffusion layer. For example, the simple transformation $x
\rightarrow (x \ll 2) \oplus (x \ggg 1)$ can be used.

Any other size $L$ can also be used as long as it has a minimal polynomial
$\Min_L(X) = X^5+X^2+1$, or any other degree 5 minimal polynomial using the
transformation described in Proposition~\ref{prop.equivalence}. Unfortunately,
not any matrix can have a minimal polynomial of degree 5. Typically, if one
wants to design a 128 bit diffusion layer, using a $16\times 16$ matrix acting
on 8 bit symbols, it will not be possible to use our solutions as an $8\times 8$
binary matrix $L$ cannot have an irreducible minimal polynomial of degree 5. A
new search using a $16\times 16$ companion matrices in $\F_{2^8}$ has to be
done, but it will be much more expensive than on $\F_{2^5}$, probably even out
of reach.

\subsection{Going Further}

The next step is $s=6$ allowing to build a $32\times 32$ matrix for a 192 bit
diffusion layer. Such a diffusion would have a branch number of 33, which is way
above anything usual in symmetric cryptography. In this sense, finding one such
matrix would be of interest, especially as it would still have a somehow compact
description (compared to a traditional $32\times 32$ MDS matrix).

The exhaustive search is however completely out of reach, with a number of
``symmetric'' companion matrices to explore around $2^{93 }$. Building such a
matrix will thus require a direct construction. The first step to getting this
direct construction is probably to understand what additional structure the
solutions we found possess.



\end{document}